\documentclass[twocolumn]{aastex62}

\graphicspath{{./}{Figures/}}
\usepackage{longtable}
\usepackage{amsmath}
\usepackage{mathtools}

\def\cgmsq {CGM$^2$~}
\def\lya{Ly$\alpha$~}

\def\NHI{$N_{\rm HI}$~}

\def\Rsp{$R_{\rm sp}$}

\usepackage{hyperref}
\hypersetup{
    pdfnewwindow=true,
    colorlinks=true,
    citecolor=cyan,
    linkcolor=blue,
    urlcolor=blue
}

\begin{document}
\title{CGM$^2$ $+$ CASBaH: The Mass Dependence of H~I Ly$\alpha$-Galaxy Clustering and the Extent of the CGM}

\correspondingauthor{Matthew C. Wilde}
\email{mwilde@uw.edu}

\author[0000-0003-1980-364X]{Matthew C. Wilde}
\affil{University of Washington, Department of Astronomy, Seattle, WA 98195, USA}

\author[0000-0003-0789-9939]{Kirill Tchernyshyov}
\affil{University of Washington, Department of Astronomy, Seattle, WA 98195, USA}

\author[0000-0002-0355-0134]{Jessica K. Werk}
\affil{University of Washington, Department of Astronomy, Seattle, WA 98195, USA}

\author[0000-0002-1218-640X]{Todd M. Tripp}
\affil{Department of Astronomy, University of Massachusetts, 710 North Pleasant Street, Amherst, MA 01003-9305, USA}

\author[0000-0002-1979-2197]{Joseph N. Burchett}
\affil{University of California, Santa Cruz; 1156 High St., Santa Cruz, CA 95064, USA}
\affil{Department of Astronomy
, New Mexico State University, PO Box 30001, MSC 4500, Las Cruces, NM 88001}

\author[0000-0002-7738-6875]{J. Xavier Prochaska}
\affil{University of California, Santa Cruz; 1156 High St., Santa Cruz, CA 95064, USA}
\affil{Kavli Institute for the Physics and Mathematics of the Universe (Kavli IPMU) The University of Tokyo; 5-1-5 Kashiwanoha, Kashiwa, 277-8583, Japan}
\affil{Division of Science, National Astronomical Observatory of Japan,2-21-1 Osawa, Mitaka, Tokyo 181-8588, Japan}

\author[0000-0002-1883-4252]{Nicolas Tejos}
\affil{Instituto de F\'{i}sica, Pontificia Universidad Cat\'{o}lica de Valpara\'{i}so, Casilla 4059, Valpara\'{i}so, Chile}

\author[0000-0001-9158-0829]{Nicolas Lehner}
\affil{Department of Physics and Astronomy, University of Notre Dame, Notre Dame, IN 46556}

\author[0000-0002-3120-7173]{Rongmon Bordoloi}
\affil{North Carolina State University, Department of Physics, Raleigh, NC 27695-8202}

\author[0000-0002-7893-1054]{John M. O'Meara}
\affil{W. M. Keck Observatory, 65-1120 Mamalahoa Hwy., Kamuela, HI 96743, USA}

\author[0000-0002-7982-412X]{Jason Tumlinson}
\affil{Space Telescope Science Institute, Baltimore, MD, USA}

\author[0000-0002-2591-3792]{J. Christopher Howk}
\affiliation{Department of Physics and Astronomy, University of Notre Dame, Notre Dame, IN 46556, USA}

\begin{abstract}
We combine datasets from the CGM$^{2}$ and CASBaH surveys to model a transition point, $R_{\rm cross}$, between circumgalactic and intergalactic media (CGM and IGM, respectively). In total, our data consist of 7244 galaxies  at z $<$ 0.5 with precisely measured spectroscopic redshifts, all having impact parameters of 0.01 $-$ 20 comoving Mpc from 28 QSO sightlines with high-resolution UV spectra that cover \ion{H}{1} Ly$\alpha$. Our best-fitting model is an exclusionary two-component model that combines a 3D absorber-galaxy cross correlation function with a simple Gaussian profile at inner radii to represent the CGM. By design, this model gives rise to a determination of $R_{\rm cross}$ as a function of galaxy stellar mass, which can be interpreted as the boundary between the CGM and IGM. For galaxies with $10^8 \leq M_{\star}/M_{\odot} \leq 10^{10.5}$, we find that $R_{\rm cross}(M_{\star}) \approx 2 \pm 0.6 R_{\rm vir}$. Additionally, we find excellent agreement between $R_{\rm cross}(M_{\star})$ and the theoretically-determined splashback radius for galaxies in this mass range. Overall, our results favor models of galaxy evolution at z $<$ 0.5 that distribute $T \approx 10^{4}$K gas to distances beyond the virial radius. 
\bigskip
\end{abstract}

\section{Introduction}

The formation and evolution of galaxies involves a complex interplay between gravitational collapse of  gas from the intergalactic medium (IGM), galaxy mergers, and feedback due to stellar evolution and active galactic nuclei (AGN) that drive gaseous outflows and change the ionization state of the galaxies' gaseous halos.  Together, these processes drive the ``cosmic baryon cycle" which takes place largely in the region of a galaxy referred to as the circumgalactic medium (CGM). Indeed, understanding the CGM is critical for developing a complete theory of galaxy evolution, as highlighted by the recent decadal survey \citep[]{decadalsurvey}. In particular, the extent of the gaseous CGM relative to the extent of the dark matter halo is a subject of great interest for models that aim to reproduce the properties of gaseous halos. 

The existence of the CGM, first predicted by \cite{bahcall69}, was initially revealed by detection of \ion{Mg}{2} and \ion{H}{1} absorption at large projected distances ($R_{\perp} > 20$ kpc) from $L*$ galaxies \citep{bergeron86, morris93, bergeron91, lanzetta95, chen05},  and subsequently traced via higher-energy metal-line transitions such as \ion{Si}{3}, \ion{C}{4} and \ion{O}{6} that are observed to correlate with galaxies and their global properties \citep[e.g.][]{tripp00, tripp08, prochaska11b, tumlinson11, werk13}. Within 0.5 $R_{\rm vir}$ of L $\sim$ L* galaxies,  the metal line incidence is found to be $60-90$ \% for a range of ionized metal species \citep{werk13}. Conversely, \cite{berg22} find an 80\% chance of finding a massive galaxy nearby to any high-metallicity absorber. The CGM of $M_{\star}>$ 10$^{8}$ $M_{\odot}$ galaxies is now well-established to be metal-enriched \citep[]{liang14, bordoloi14, prochaska17, berg22}, and to extend to at least 1 $R_{\rm vir}$, and very likely beyond it \citep{wakker09, burchett15, finn16, wilde21, borthakur22}.

Generally, hydrodynamical simulations of galaxy evolution, which exhibit complex interactions between gravitational collapse from the cosmological large scale structure and subsequent feedback from supernovae and AGN-driven winds that heat and enrich the CGM and IGM (EAGLE, \citealt{schaye15}; IllustrisTNG, \citealt{pillepich18}; SIMBA, \citealt{dave19}; and CAMELS, \citealt{villaescusa-navarro22}), are consistent with the range of observations of the CGM in absorption. Yet these models still rely on simplistic implementations of the ``sub-grid" physics in order to model entire galaxies \citep[e.g.][]{ford13, hummels13}, and physical properties of the CGM are dependent on the simulation resolution \citep[][]{hummels19,peeples19}. More sensitive observations of the CGM, including the ability to detect the diffuse gas in emission, are needed both to break degeneracies in these models, e.g., between heating and cooling mechanisms, and to develop a flexible parametric model of the CGM \citep{singh21}. 

The two-point correlation function between \ion{H}{1} absorption along QSO sightlines and galaxies has proven to be an essential tool to understand the connection of galaxies to the IGM \citep[e.g.][]{morris93, chen05, ryan-weber06, prochaska11b, tejos14, prochaska19}. The primary advantages of leveraging the clustering of these two entities over one-to-one association analyses is that it provides results for large scales (1-10 Mpc) as well as the relatively smaller scales where the baryonic processes associated with the CGM play out, and the correlation function statistically characterizes absorber-galaxy relationships when multiple galaxies are close to the sightline and a one-to-one assignment is ambiguous. Since \ion{H}{1} traces both enriched material from galaxies as well as primordial accretion from the IGM, observations of the CGM, IGM, and galaxies in the same volume are fundamental to both testing  the predictions of galaxy evolution models and providing a means to differentiate between them \citep[e.g.][]{fumagalli11, oppenheimer12, stinson12,  ford13, hummels13, butsky20, singh21}.  

Understanding the physical profile and size of the CGM sheds light on the non-linear processes of galaxy formation: on what spatial scale(s) do virialization, accretion, and feedback transform these galactic atmospheres? Astronomers have long used some version of the virial radius as an estimator for the size of galaxy halos, but this estimate is somewhat arbitrary and is based on the distibution of unobservable dark matter. By observing the radial gas profile around galaxies out to large scales, we can effectively map the gaseous halo, which in turn constrains the physics of galaxy-scale feedback processes. Observationally determining the galactic atmosphere's extent has additional implications for constraining galaxy evolution and assembly models. For example, the galaxy baryon and metal budgets require a scale to integrate the total mass \citep[e.g.][]{peeples14, werk14}. Furthermore, the gaseous halo likely plays an important role in the quenching of dwarf satellite galaxies as they become stripped by ram-pressure in a low-density CGM \citep[]{putman21}, and it is useful to constrain where this occurs, i.e., the extent of the CGM, and how this depends on central galaxy mass.

The presence of \ion{H}{1} absorption beyond the virial radius is now widely accepted for a range of galaxy stellar masses \citep[e.g.][]{prochaska11b, tejos12, tejos14, wilde21, bouma21, borthakur22}. In \cite{wilde21} (Paper I) we found an empirical relation between galaxy stellar mass and the extent of the CGM as indicated by \ion{H}{1} covering fractions. For galaxies with stellar masses $10^{8} < M_{\star}/M_{\odot} < 10^{11.5}$, we found that the CGM extends to two times the virial radius.   In this paper, we focus on the functional forms of the mass dependence of the \ion{H}{1}-traced CGM using a power-law model similar to the 2-halo correlation function. We also investigate other two-component models that differentiate the CGM from the IGM. We combine the \cgmsq Survey, which focuses on sightlines at low galaxy impact parameters ($<1$ Mpc), with the \emph{COS Absorption Survey of Baryon Harbors} (CASBaH) that probes larger spatial scales ($<20$ Mpc). In doing so, we greatly increase the absorber-galaxy sample from 543 spectroscopically-confirmed absorber-galaxy pairs to 7244 pairs spanning $0.003 < z < 0.48$. Our goal is to provide the most reliable constraints to date on the spatial extent of the CGM as traced by \ion{H}{1} absorption. 

The paper is structured as follows: In Section \ref{section:data}, we briefly review each of the galaxy-absorber surveys and discuss their combined properties. In Section \ref{section:models}, we introduce two models of the \ion{H}{1}-galaxy correlation functions and cover our main results in Section \ref{section:results}. We compare our results with simulations and previous results and discuss their implications for galaxy evolution models in Section \ref{section:discussion}. Finally, we summarize our results in Section \ref{section:summary}.

\section{Data - Combining \cgmsq and CASBaH}\label{section:data}

\begin{figure}$
\begin{array}{c}
\includegraphics[scale=0.6]{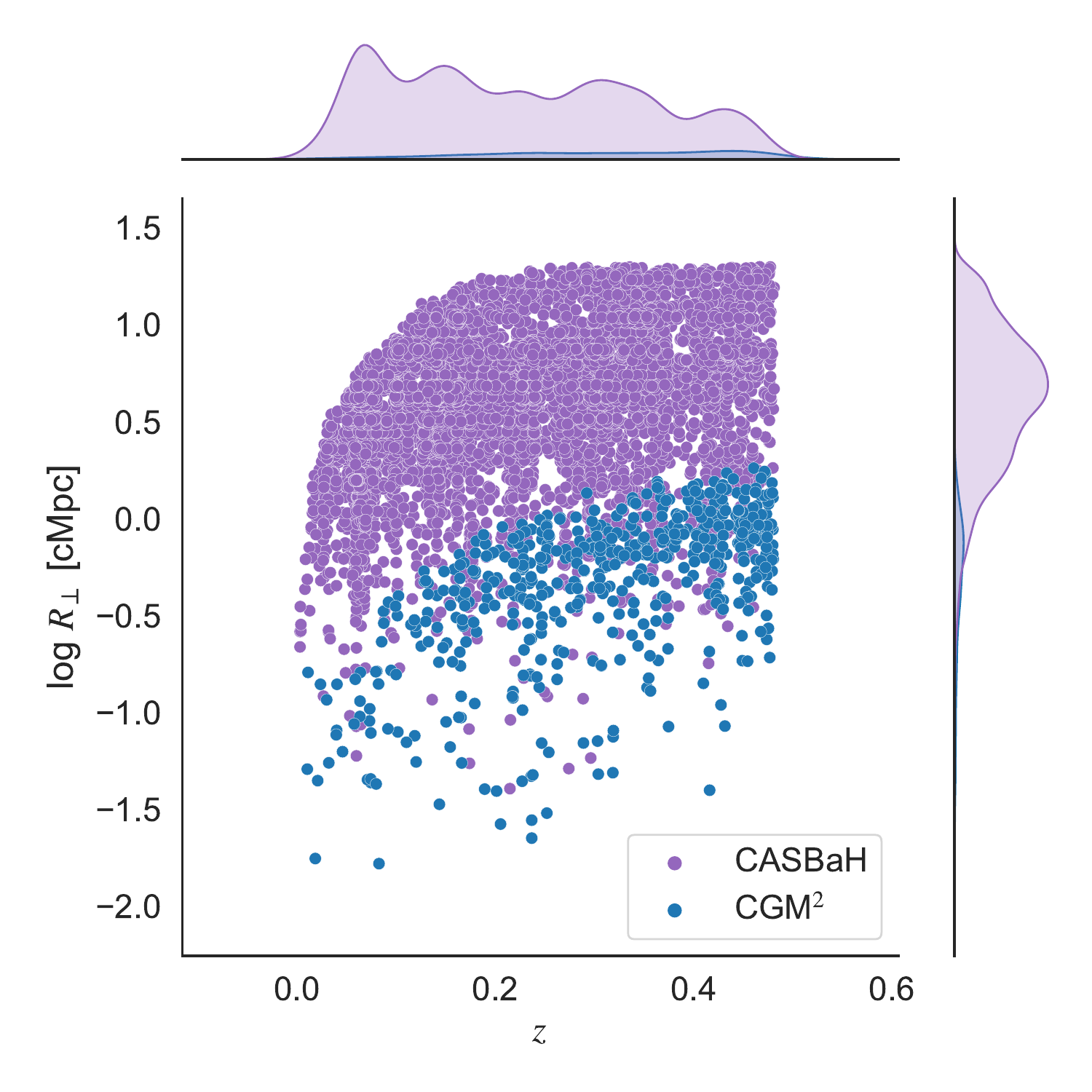} \\
\includegraphics[scale=0.6]{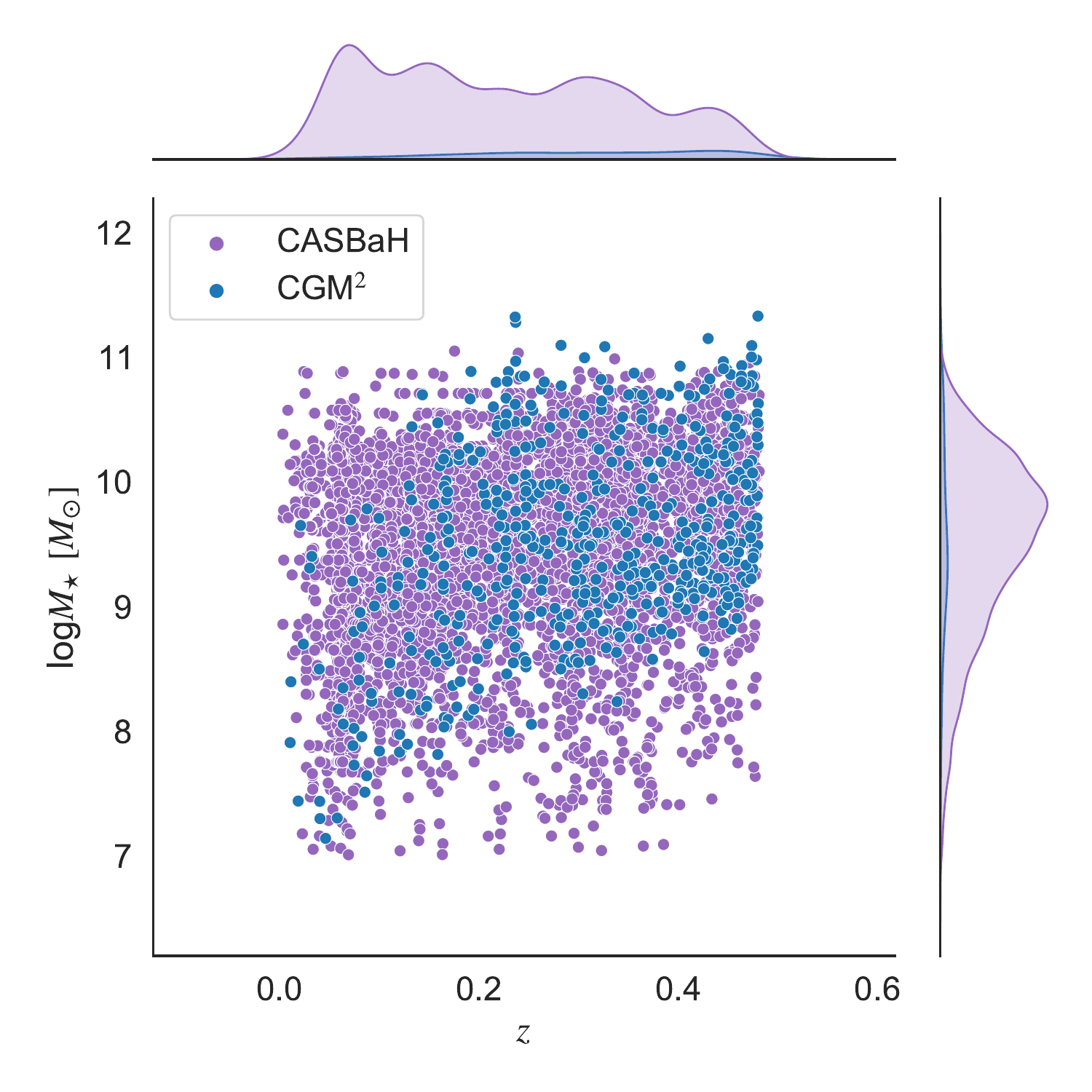} \\
\end{array}$
\vspace{+5pt}
\caption{\textbf{Top}: Distribution of the combined \cgmsq (blue dots) and CASBaH (purple dots) data sets in both logarithmic impact parameter, and redshift. The data are roughly uniform in redshift space but we can see the relative contributions of the data sets in impact parameter space; \cgmsq is highly concentrated at lower impact parameters while CASBaH explores much greater impact parameters. \textbf{Bottom}: Galaxy stellar mass distribution as a function of redshift for the two data sets.
}
\label{fig:data}
\end{figure}

    Both surveys feature far-ultraviolet spectroscopy of QSOs with \emph{HST}, using both the \emph{Cosmic Origins Spectrograph}  \citep[COS,][]{green12} and the \emph{Space Telescope Imaging Spectrograph} \citep[STIS,][]{woodgate98}, and deep, ground-based optical spectroscopy of foreground galaxies in the QSO fields. CASBaH is well suited to the study of the interface between the CGM and the IGM, at scales $\gtrsim 1$ Mpc. \cgmsq provides a relatively more complete mapping of the inner CGM at scales $\lesssim 1$ Mpc. By combining \cgmsq and CASBaH data, we leverage the strengths of each survey, as described below.  Figure (\ref{fig:data}) shows the distributions of galaxy stellar masses and impact parameters versus redshift from both surveys out to z $= 0.5$. Together, the surveys allow us to probe the CGM as it transitions into the IGM for a large sample of galaxies. 
       
    \subsection{\cgmsq} 
    
    The \cgmsq survey, first presented in \cite{wilde21}, includes precise spectroscopic redshifts and bulk galaxy properties (e.g.~stellar masses, M$_*$,  and star formation rates, SFR) from a combination of Gemini GMOS spectra and deep, broadband photometry for $\sim$1000 galaxies in the foreground of 22 QSOs, each with S/N $\approx$10 HST/COS G130M$+$G160M spectra. By matching galaxy and absorber redshifts in $\pm$500 km s$^{-1}$ windows, the \cgmsq survey is ultimately a large collection of measurements pertaining to the CGM of z $<$ 1 galaxies over a wide range of stellar masses, 10$^{8}$ $\lesssim M_{\star}/M_{\odot} \lesssim 10^{11.5}$. The data acquisition and analysis are explained in detail in \cite{wilde21}. Here we present a brief overview of the survey data relevant to the present analysis. 
    
    The \cgmsq galaxy spectra were obtained using Gemini-GMOS spectrographs on the twin Gemini North and South telescopes \citep[]{hook04, gmoss16}. Galaxy redshifts were inferred from the template fitting code, Redrock\footnote{https://github.com/desihub/redrock} (v0.14) and manually inspected with \texttt{VETRR}\footnote{https://github.com/mattcwilde/vetrr}. The typical statistical uncertainly of our redshifts is $\sigma_z \sim 50$-$100$ km s$^{-1}$ ($z \simeq 0.00016$-$0.00030$). Photometry of the \cgmsq galaxy catalog was obtained from the Gemini-GMOS pre-imaging in $g$ and $i$ bands as well as all available bands from DESI Legacy Imaging Surveys Data Release 8 (DR8) \citep{dey19}, WISE \citep{cutri13}, Pan-STARRS Data Release 2 \citep{chambers16}, and SDSS DR14 \citep{aboltathi18}. 
    
    The 22 QSOs included in the \cgmsq survey have \emph{HST}/COS spectra selected from the COS-Halos (GO11598, GO13033; \citealp{tumlinson13}) and COS-Dwarfs (GO12248;
    \citealp{bordoloi14}) surveys. In general, the \cgmsq QSO targets have  $z_{\rm QSO} >$ 0.6 and available HST imaging, which permits detailed analysis of absorption-hosting galaxies with $z < 0.5$. All COS spectra include both the G130M and G160M gratings, and have a S/N $\simeq 8-12$ per resolution element (FWHM $\simeq$ 16-18 km s$^{-1}$) or better over $1150$-$1800$~\AA. The COS data and their reduction are presented in detail in \cite{tumlinson13} and \cite{bordoloi14} and follows the same method used by \cite{tripp11}, \cite{meiring11}, \cite{tumlinson11} and \cite{thom12}. 
    
    \subsection{CASBaH}
   
    The CASBaH program was designed to take advantage of the multitude of resonance transitions at rest-frame wavelengths $<$ 912 \AA\ to probe the physical conditions, metallicity, and physics of the multiphase CGM.  A wide variety of elements and ionization stages have resonance lines only at $\lambda <$ 912 \AA~\citep[see, e.g.,][]{verner94}, so observations of this wavelength range provide new diagnostics and precise constraints using banks of adjacent ions such as \textsc{N~i} through \textsc{N~v}, \textsc{O~i} through \textsc{O~vi}, and Ne~\textsc{ii} through Ne~\textsc{viii} \citep[see][for examples of lines detected by CASBaH]{tripp11}. The Ne~\textsc{viii} 770.4, 780.3 \AA\ doublet has received particular attention as a probe of warm-hot gas at $\approx 10^{5} - 10^{6}$ K \citep[e.g.,][]{savage05,burchett19,wijers20}.  In many contexts such as the Milky Way interstellar medium, these lines are inaccessible because they are blocked by the \textsc{H~i} Lyman limit. CASBaH overcomes this limitation by observing QSO absorbers with sufficient redshift to bring the lines into the observable band of HST.
    
    The motivation and design of the CASBaH program is summarized in section 1 of \citet{haislmaier21}, and the CASBaH galaxy redshift survey is presented in \citet{prochaska19}. Briefly, CASBaH obtained  both {\it HST}/COS and {\it HST}/STIS spectra of nine QSOs at 0.92 $< z_{\rm QSO} <$ 1.48, with two primary selection criteria. First, since some of the most important target lines (e.g., Ne~\textsc{viii}) are weak, the QSOs were required to be UV-bright so that good signal-to-noise and sensitivity to weak lines would be attained. Second, the targets were required to have $z_{\rm QSO} > 0.9$ to provide a total redshift path that is sufficient to accumulate a statistically useful sample of absorbers of interest. No considerations were given to known foreground galaxies or absorbers, so the targets were not selected in a way that would favor particular types of foreground absorbers or galaxies, except that sightlines with known black Lyman limits at $\lambda_{\rm ob} > 1150$ \AA\ were excluded to avoid using HST time on sightlines that would not contribute useful pathlengths to the samples  \citep[see][]{burchett19}.  The CASBaH UV spectra were reduced in the same way as the CGM$^{2}$  data.
    
    The CASBaH galaxy-redshift survey \citep{prochaska19} measured thousands of redshifts in the fields of seven of the CASBaH QSOs using the Keck DEIMOS and MMT Hectospec spectrographs, with typical redshift uncertainties of $\approx$ 30 km s$^{-1}$. The survey used a wedding-cake strategy with the Hectospec covering galaxies in the $\approx 1^{\circ}$ fields centered on the QSOs and the DEIMOS survey providing a deeper survey with a smaller field of view (81.5 arcmin$^2$) \citep[see][]{prochaska19}. Using the CASBaH galaxy database, supplemented with data from public surveys such SDSS, we selected a sample of 6701 galaxies with spectroscopic redshifts $z < 0.481$ and comoving impact parameters less than 13 cMpc, appropriate for the \ion{H}{1}\ analysis presented here. 
    
    \subsection{Synergy of CGM$^{2}$ + CASBaH}
    
    The CASBaH and CGM$^{2}$ surveys have complementary designs. On the one hand, CGM$^{2}$ is built on COS-Halos and thus favors at least one $L*$ galaxy close to the sightline. CGM$^{2}$ also covers a smaller FOV. On the other hand, CASBaH is a blind survey that covers a larger FOV.  Consequently, CASBaH provides more information about galaxies and large-scale structures at larger impact parameters, but as a blind survey, it is cross-section weighted in favor of galaxies at larger impact parameters. Also, since CASBaH avoided sightlines with black Lyman limits in the HST band (i.e., at $\lambda_{\rm{ob}} \geq 1150$ \AA ), it will not include galaxies at $z_{\rm gal} > 0.26$ that harbor absorbers with $N$(\textsc{H~i}) $\gtrsim 10^{17}$ cm$^{-2}$.  Thus, CGM$^{2}$ probes the inner CGM including higher $N$(\textsc{H~i}) absorbers, while CASBaH complements CGM$^{2}$ by adding very large samples of galaxies and structures at larger distances.

    \subsection{Galaxy Properties}
    To estimate the galaxy properties for both surveys, we used CIGALE \citep[]{cigale11, boquien19} to fit the spectral energy distribution (SED) and retrieve stellar mass and star formation rates (SFR). We used the \cite{bruzual03} stellar population models, assuming a \cite{chabrier03} initial mass function (IMF). We chose a grid of metallicities ranging from $0.001$-$2.5 Z_{\odot}$. A delayed star formation history (SFH) model was employed with an exponential burst. The e-folding time of the main stellar population models ranged from 0.1-8~Gyr. 
    We varied the age of the oldest stars in the galaxy from 2-12~Gyr. We included an optional late burst with an e-folding time of 50~Myr and an age of 20 Myr. The burst mass fraction varied from 0.0 or 0.1 to turn this feature on or off. Nebular emission and reprocessed dust models \citep{dale14} were also included with the default values. The dust models have slopes ranging from $1-2.5$ and the nebular models include no active galactic nuclei.

    We employed the \cite{calzetti94} dust attenuation law, but we also included a ``bump" in the UV \citep[see discussion in][]{prochaska19} at 217.5~nm with a FWHM of 35.6~nm. The bump amplitude is set at 1.3 and the power law slope is -0.13 \citep{lofaro17}. We varied the color excess of the stellar continuum from the young population, E(B-V), from 0.12-1.98. Finally, we used a reduction factor of 0.44 to the color excess for the old population compared to the young stars. 
    
    CIGALE then provides us with Bayesian estimates for the stellar mass and SFR for each galaxy in the combined catalog. In order to calculate the virial radius we used the abundance matching method of \cite{moster13} with the modifications used in \cite{burchett16}. We adopt the convention of using $R_{\rm vir}$ = $R_{200m}$, the radius within which the average mass density is 200 times the mean matter density of the universe, as the virial radius ($R_{\rm vir}$) of a galaxy halo.
    
    \subsection{Combining the \cgmsq and CASBaH Surveys}
    In order to combine the surveys, we modified both catalogs to ensure the same matching criteria between galaxies and absorbers. In the original \cgmsq survey, we measured the $2\sigma$ upper limit on absorption within $\delta v = \pm 30$ km s$^{-1}$ of the galaxies redshift using the normalized error of the quasar flux when no absorption system was found within our $|\delta v| < 500$ km s$^{-1}$ window. In order to match the CASBaH survey, we adjusted this to a $3\sigma$ upper limit. This did not change our results in a meaningful way. The original CASBaH survey used a velocity window of $|\delta v| < 400$ km s$^{-1}$ to match the galaxies to absorption systems. We adjusted the window for this work to $|\delta v| < 500$ km s$^{-1}$ to match the \cgmsq survey. As in Paper I, we restrict our \ion{H}{1} measurements to those less than $z < 0.481$ since at this redshift, the Lyman-$\alpha$ line redshifts out of the G160 grating band, and thus we are only sensitive to higher order transitions at higher redshifts. 
    
    Having made these two small changes to each survey, both could be combined to give us a total survey that includes 7244 galaxies spanning $\sim 0.01-8$ comoving Mpc in impact parameter around 28 QSO sightlines. The distributions of impact parameter, redshift, and stellar mass are shown in Figure \ref{fig:data}. In this paper, we will focus on galaxies with $8 < \log M_{\star}/M_{\odot} < 10.5$, a stellar mass range with good coverage in both surveys, which trims our galaxy sample to 6136 galaxies from CASBaH and 453 galaxies from CGM$^{2}$ for a total sample of 6589 absorber-galaxy pairs. The number of absorber-galaxy pairs is summarized in Table \ref{tab:galdata}.

\begin{deluxetable*}{cccccc}
\tablecaption{Number of Absorber-Galay Pairs \label{tab:galdata}}
\tablehead{\colhead{Survey} & \colhead{$10^{7-11.3} M_*/M_{\odot}$} & \colhead{$10^{8-10.5} M_*/M_{\odot}$} & \colhead{$10^{8-9} M_*/M_{\odot}$} & \colhead{$10^{9-10} M_*/M_{\odot}$} & \colhead{$10^{10-10.5} M_*/M_{\odot}$}
}
\colnumbers
\startdata
CGM$^2$ &    543 &  453 &  103 &  271 &   79 \\
CASBaH  &   6701 & 6136 & 1265 & 3545 & 1326 \\
Total   &   7244 & 6589 & 1368 & 3816 & 1405 \\
\enddata
\tablecomments{Summary of absorber-galaxy pairs used in this manuscript. (1) The number of absorber-galaxy pairs in each survey and total of the combined surveys; (2) the number of absorber-galaxy pairs in the entire mass range; (3) the mass range used to perfom the model fitting; (4, 5, 6) the number of absorber-galaxy pairs within each mass bin used for model verification.}
\end{deluxetable*}

\section{Modeling Absorber-Galaxy Clustering}\label{section:models}

We model the CGM using an absorber-galaxy cross-correlation analysis. This technique is based on modeling the covering fraction, $f_c$, as a binomial probability distribution of detections. To ensure high completeness in the absorber sample, based on the S/N of the data, we require a total column density \NHI $\geq 10^{14}$ cm$^{-2}$ to consider the sightline to have a ``detection''. Likewise, a non-detection is the case where we do not detect gas above this threshold. The models used here are based on the models employed in Paper I, which was inspired by the model developed by \cite{hennawi07} and  \cite{prochaska19}. A more detailed explanation can be found in those three papers. In Paper I, we found a mass dependence of the extent of the CGM based on dividing the data into three mass bins. In this work, we wish to quantify the mass dependence of the clustering as well as determine the redshift dependence given our data. 

\subsection{Single Power-Law Model}
\label{sec:single-pl}

\begin{figure*} 
\epsscale{0.85}
\plotone{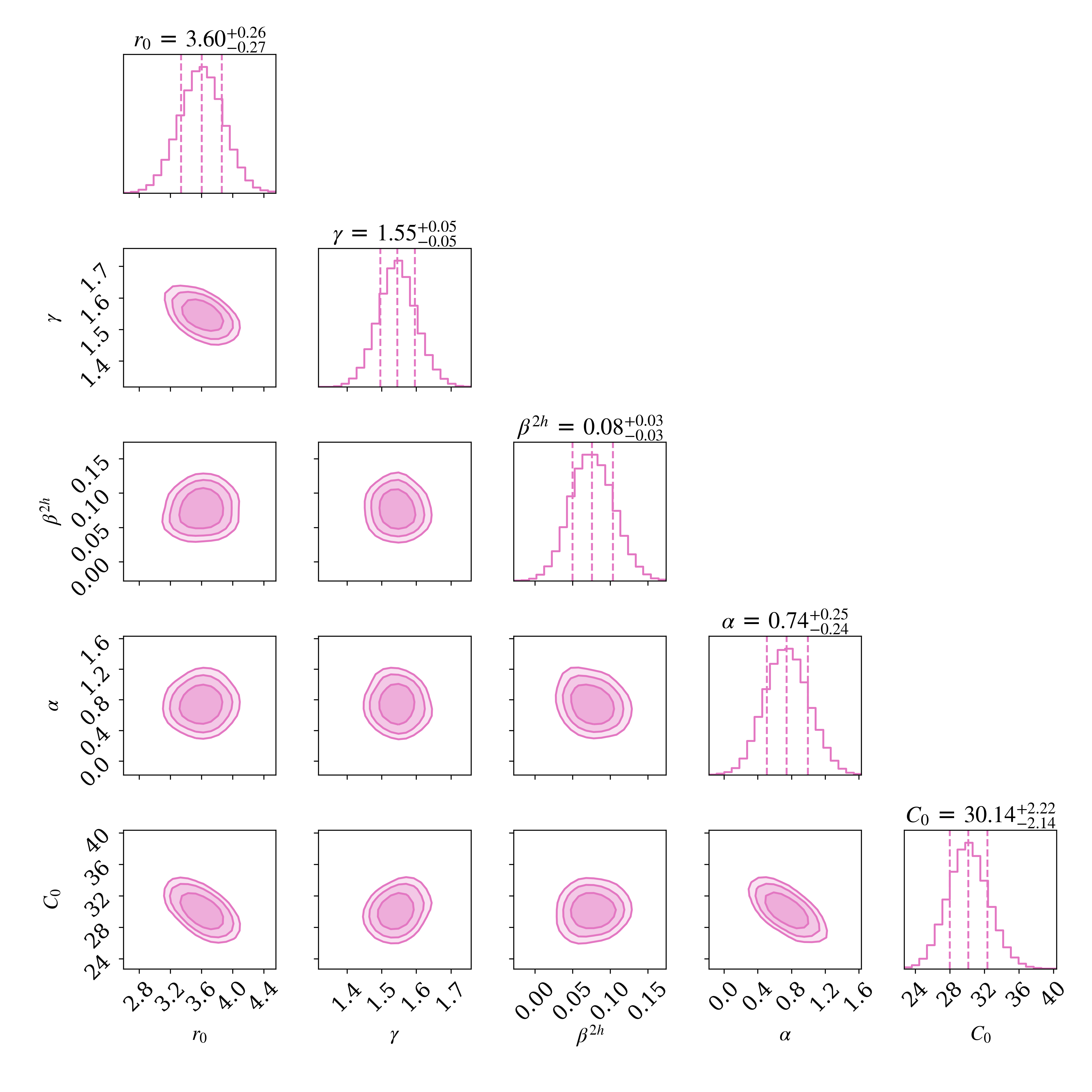}
\caption{Corner plots showing the posterior parameter probabilities for the parameters in the  single power-law clustering model. We find a non-zero, positive mass dependence term in the two-halo absorber-galaxy clustering, $\beta^{2h}$. \label{fig:corner_2h}}
\end{figure*}

The single power-law model consists of two terms: the base rate of detection due to the random incidence of absorbers greater than this threshold and an excess above this base rate due to the clustering of galaxy-absorber pairs.

Much like \cite{prochaska19}, we define the 3D absorber-galaxy cross-correlation function, $\xi_{ag}(r)$ as

\begin{equation} \label{eq:xi}
    \xi_{ag}(r) = \left( \frac{r}{r_0}\right)^{-\gamma}.
\end{equation} 

To model the galaxy mass dependence of the clustering, we add a new mass dependence to the clustering scale, $r_0$, 
\begin{equation} \label{eq:eq_r0}
    r_{0,m}(m) = r_{0}\left(\frac{M_{\star}}{M_0}\right)^{\beta}.
\end{equation}

As before, we examine the projected 2-D correlation function, which is obtained by integrating the 3-D correlation function over the line of sight
\begin{equation} \label{eq:chiperp}
    \chi_{\perp}(r) = \frac{1}{\Delta r_{\parallel}} \int_{r_{\parallel}} \xi_{ag}(\sqrt{r_{\parallel}^2 + r_{\perp}^2}~) dr_{\parallel}, 
\end{equation}

\noindent where $r_{\parallel}$ is the line-of-sight distance,  $r_{\perp}$ is the transverse distance, and $\Delta r_{\parallel}$ is the size of the redshift window. 

For simplicity of notation, $r$ is equivalent to $r_{\perp}$ in the following analysis. 

In the following definitions, we label the single power law clustering terms ``2-halo," as the galaxy clustering method we adopt here describes the clustering of separate dark matter halos. This approach distinguishes the ``two-halo" only method from the two-component model we develop later in this manuscript. 

In order to model $f_c$, we assume that the number of detected absorbers above the column-density threshold has a Poisson distribution. We consider two cases: (1) one or more absorbers detected, and (2) the case where no absorbers are detected. In this framework the probability of seeing no absorbers is 

\begin{equation}\label{eq:pmiss}
    P^{\rm miss} = \frac{\lambda^0\exp(-\lambda)}{0!}
\end{equation}

\noindent where we denote the rate of incidence (see below) as $\lambda$. The probability of finding one or more absorbers is just the complement of Equation \ref{eq:pmiss},

\begin{equation}
    f_c = 1 - P^{\rm miss}.
\end{equation}

We model the rate of absorber incidence as the projected correlation function, the 2-halo term, as the excess over the probability of intersecting an absorber with \NHI $> 10^{14}$ cm$^{-2}$ in the redshift window, 

\begin{equation}
    \lambda = (1 + \chi_{\perp}^{2h})~\langle d\mathcal{N}/dz\rangle \delta z,   
\end{equation}

\noindent where $\langle dN/dz\rangle$ is the base rate of detection due to the random incidence of absorbers greater than this threshold and $delta z$ is the line-of-sight redshift window. 

In addition to parameterizing the mass dependence as in Equation (\ref{eq:eq_r0}), we also  parameterize the redshift dependence of $\langle dN/dz\rangle$ as follows: 

\begin{equation} \label{eq:eq_dndz}
    \frac {d\mathcal{N}(\rm N_{\rm HI} \geq \rm N_{\rm HI}^{14} ,z)}{dz} =
    C_0 (1+z)^{\alpha},
\end{equation}

\noindent where $N_{\rm HI}^{14}$ denotes absorbers with column densities of $10^{14}$ cm$^{-2}$, $C_0$ is the random rate of incidence at $z = 0$, and $\delta z$ is the redshift window. We adopt a redshift window to be $\pm 500$ km s$^{-1}$ in velocity units.

Thus, we have a rate of incidence of the form 

\begin{equation}
\begin{multlined}
    \lambda = (1 + [\chi_{\perp}^{2h}(r, m|r^{2h}_0, \gamma^{2h}, \beta^{2h})])~\langle d\mathcal{N}(z | C_0, \alpha)/dz\rangle~\delta z.
    \end{multlined}
\end{equation}

Finally, we construct the likelihood function,
\begin{equation} \label{eq:eq_likelihood}
    \mathcal{L} = \prod_{i} P^{\rm hit}(r_{i}, z_{i}, m_{i} | \theta) \prod_{j} P^{\rm miss}(r_{j}, z_{j}, m_{j}| \theta), 
\end{equation}

\noindent where $\theta = [r^{2h}_0, \gamma^{2h}, 
\beta^{2h}, C_0, \alpha]$.

In constructing our Bayesian model, we must choose priors. For the single power law parameters, we chose the priors based on the results of cross-correlation analysis by \cite{tejos14} except for our new mass dependent term, $\beta^{2h}$, which was motivated by physical arguments:
    
\begin{itemize}
    \item $r^{2h}_0 \sim \mathcal{N}(\mu = 3.2, \sigma = 0.3),~ r^{2h}_0 > 0$
    \item $\gamma^{2h} \sim \mathcal{N}(\mu = 1.7, \sigma = 0.1)$, $\gamma^{2h} > 0$
    \item $\beta^{2h} > 0$,
\end{itemize}

\noindent where $\mathcal{N}$ is the normal distribution with mean $\mu$ and variance $\sigma^2$.

The priors for the redshift dependence were chosen based on the findings in \cite{kim21}:

\begin{itemize}
    \item $C_0 \sim \rm{Lognormal}(\mu = 1.25, \sigma = 0.11)$ , $C_0 > 0$
    \item $\alpha \sim \mathcal{N}(\mu = 0.97, \sigma = 0.87)$ , $-3 < \alpha < 3$
\end{itemize}

\noindent We note that we chose to use the more recent results of \cite{kim21} in modeling the redshift evolution instead of that from \cite{danforth16}, as were used in Paper I.

As in Paper I, we apply the Bayesian Markov Chain Monte Carlo (MCMC) sampler \texttt{emcee} \citep{emcee} to generate samples from the posterior probability distribution function to estimate the parameters of interest and their distributions, using Equation (\ref{eq:eq_likelihood}) and the priors described above.

In Figure \ref{fig:corner_2h}, we show the posterior distributions of our single power-law model with $M_0 = 10^{9.5} M_{\odot}$. These were fit only to data with $8 < \log M_{\star}/M_{\odot} < 10.5$, as above this range there is a change in the virial radius due to the $M_{\star} - M_{\rm halo}$ relation from abundance matching \citep{moster13}.  Below this mass range we find a very flat covering fraction profile, which does not show a clustering signal.

\subsection{Two-component Models}\label{section:twocompmodel}
The single power-law model used in galaxy-galaxy clustering and adapted above to model the galaxy-absorber clustering makes no assumption of a CGM or overlapping (in projection) gaseous halos. However, the existence of the CGM is now well-established \citep{tumlinson17}. In particular, the trends of ionized metal species with impact parameter around L* and sub-L* galaxies from $z = 0-3.5$ distinctly show that metal-enriched gaseous atmospheres are a fundamental component of galaxies \citep[e.g.][]{werk13, lehner14, bordoloi14, borthakur15, rudie19}. In the following section, we therefore assume the existence of the CGM and use a simple Gaussian profile to model the excess clustering signal due to the presence of the CGM. In addition, we investigated several other functional forms of the CGM component, which we describe in \S\ref{section:othermodels}. We find that the particular functional form of this component has little impact on the results. 

\subsubsection{The Gaussian CGM Two-Component Model}
\label{section:gaussianmodel}
\begin{figure}[ht!] 
\includegraphics[width=\linewidth]{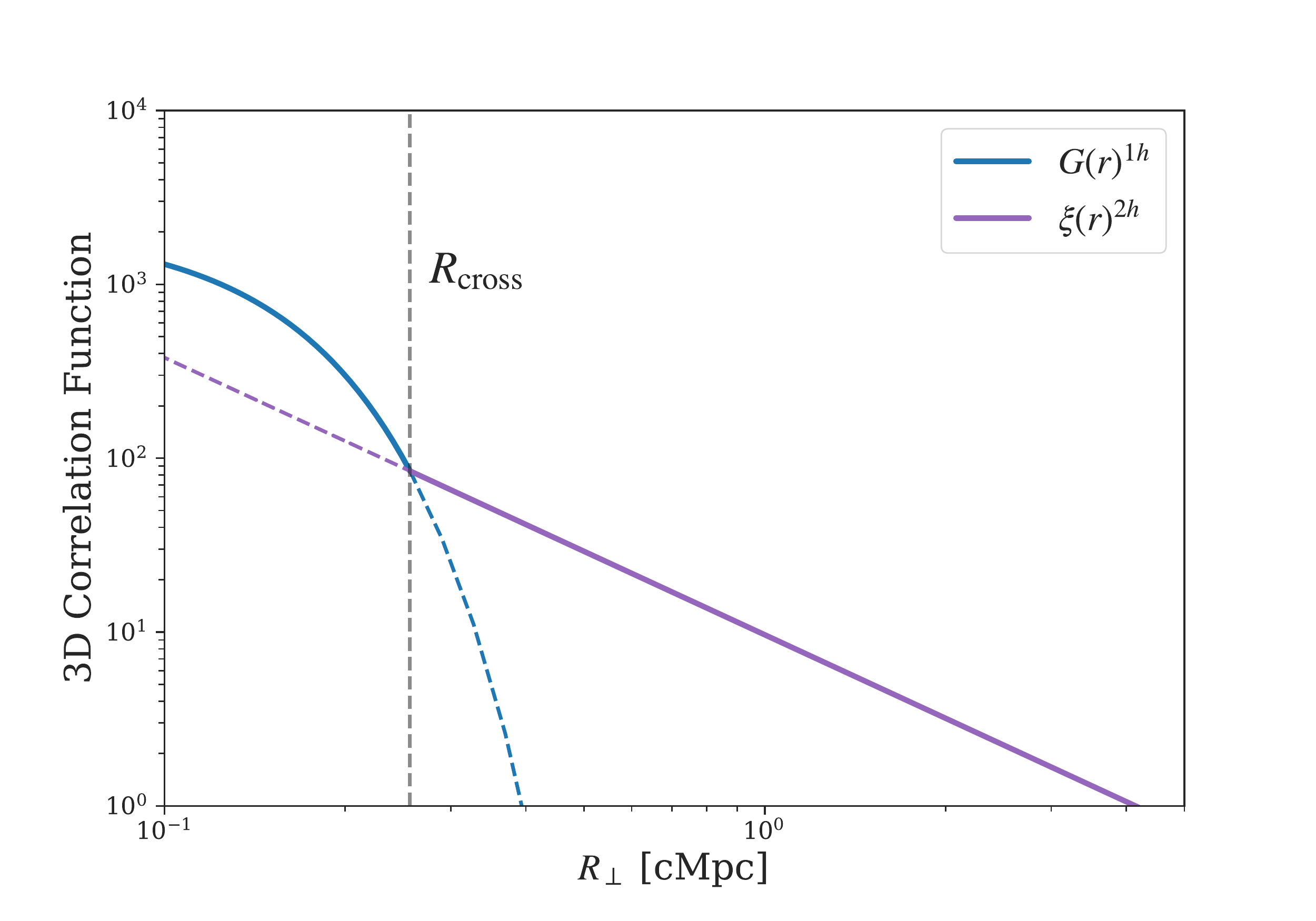}
\caption{A schematic depiction of our two-component exclusion model and the determination of $R_{\rm cross}$. The 2-halo component cuts off interior to $R_{\rm cross}$. \label{fig:schematic}}
\end{figure}

We now add a third term to the detection rate: a Gaussian 1-halo component. The detection rate now consists of a baseline random incidence rate, an enhancement due to large-scale absorber-galaxy clustering, and an additional enhancement due to the CGM. We employ an exclusion model where the contribution from the 2-halo term terminates at the distance it reaches the 1-halo component. This scheme, shown in Figure \ref{fig:schematic}, also allows us to determine a natural estimate of the extent of the CGM: the crossing point of the 1-and 2-halo components. More explicitly, within some radius, the galaxy has a CGM that we define as the gas of that galaxy and any other satellite galaxies within its halo.  Our formalism then defines the $R_{\rm cross}$ where this CGM component exceeds the 2-halo. 

The model is similar to that single power-law we introduced before with a few key differences. We introduce a Gaussian one-halo term defined as:

\begin{equation} \label{gauss}
    G(r)^{1h} = A e^{-(r/\sigma)^2}.
\end{equation}

Where the two models intersect, $R_{\rm cross}$, we can solve for $\sigma$ as

\begin{equation}
    \sigma = \sqrt{\frac{1}{2}\frac{R_{\rm cross}^2}{\ln(A) + \gamma \ln(R_{\rm cross}/r_0)}}.
\end{equation}

It should be noted that $R_{\rm cross}$ here is the 3-D distance and not the projected distance. In order to characterize the mass dependence of $R_{\rm cross}$ we define 
\begin{equation}
    R_{\rm cross} = R_{cross,0}\left(\frac{M_{\star}}{M_0}\right)^{\beta^{1h}},
\end{equation}
where $R_{cross,0}$\ is the 1-halo term extent for a galaxy at the fixed pivot mass $M_0$. The galaxy mass dependence of $\sigma$ includes contributions from the mass dependencies of $R_{cross}$ and $r_0$. 

This parameterization allows us to compare the mass dependence of the 1-halo term, $\beta^{1h}$ with that of the 2-halo term, $\beta^{2h}$.

In order to solve for the projected clustering signal, $\xi$, we first make some definitions to ease the notation. We use $s = r_{\parallel}$ in the remainder of the analysis. The integration is performed over different portions of the line of sight distance, s, corresponding to the 1 and 2-halo components. We define the line of sight crossing point $s_{\rm cross}$ as

\begin{equation}
    s_{\rm cross} = \sqrt{\max(R_{\rm cross}^2 - r_{\perp}^2, 0)},
\end{equation}

\noindent and we can then integrate Equation \ref{gauss} to $s_{\rm eval} = \min(s_{\rm cross}, s_{\rm max})$, where $s_{\rm max}$ is the maximum interval we wish to integrate over, which in our case is $[-500, 500]$ km s$^{-1}$. Thus we have

\begin{equation}
    \chi(r_{\perp}) \propto 2 \int_{0}^{s_{\rm eval}} G(r_{\perp},s)^{1h} ds  + 2\int_{s_{\rm eval}}^{s_{\rm max}} \xi(r_{\perp}, s)^{2h} ds 
\end{equation}

\noindent where the factor of 2 comes from the fact that both components are symmetric. Here we integrate the one-halo component over the more nearby regime out to $s_{\rm eval}$ and only integrate the 2-halo term beyond $s_{\rm eval}$ out to the maximum line of sight distance, thus excluding the regimes in which the models do not apply. For the two-component model, we choose fairly weak priors on unknown parameters based on physical arguments while following the same priors as described above for the parameters in the single power-law model:

\begin{itemize}
    \item $\beta^{1h} > -3$
    \item $A > 0$
    \item $R_{\rm cross} > 0$
\end{itemize}

\begin{figure*} 
\epsscale{1.1}
\plotone{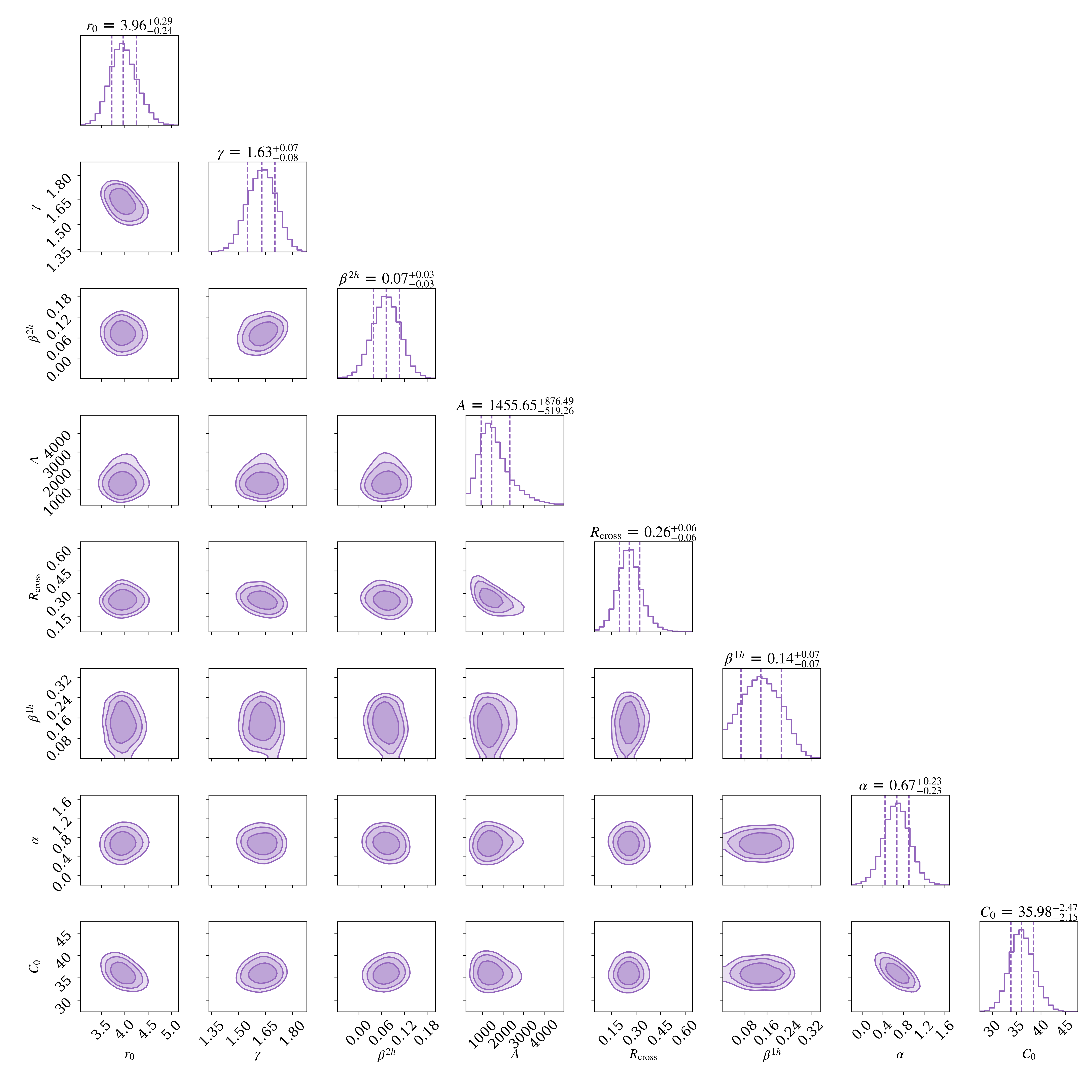}
\caption{Posterior probabilities for the parameters in the  two-component clustering model. We again recover a non-zero, positive mass dependence term in the two-halo absorber-galaxy clustering, $\beta^{2h}$ but find an even stronger one-halo CGM clustering mass dependence $\beta^{1h} \simeq 0.14 \pm 0.07$. \label{fig:corner_1h}}
\end{figure*}

We can then follow the same MCMC fitting procedure described above to determine the posteriors for the parameters in this model as well as the crossing radius, $R_{\rm cross}$. These are shown in Figure \ref{fig:corner_1h}. As before, we only fit data with $8 < \log M_{\star}/M_{\odot} < 10.5$ and use $M_0 = 10^{9.5} M_{\odot}$.

\begin{figure*}[ht!] 
\includegraphics[width=\linewidth]{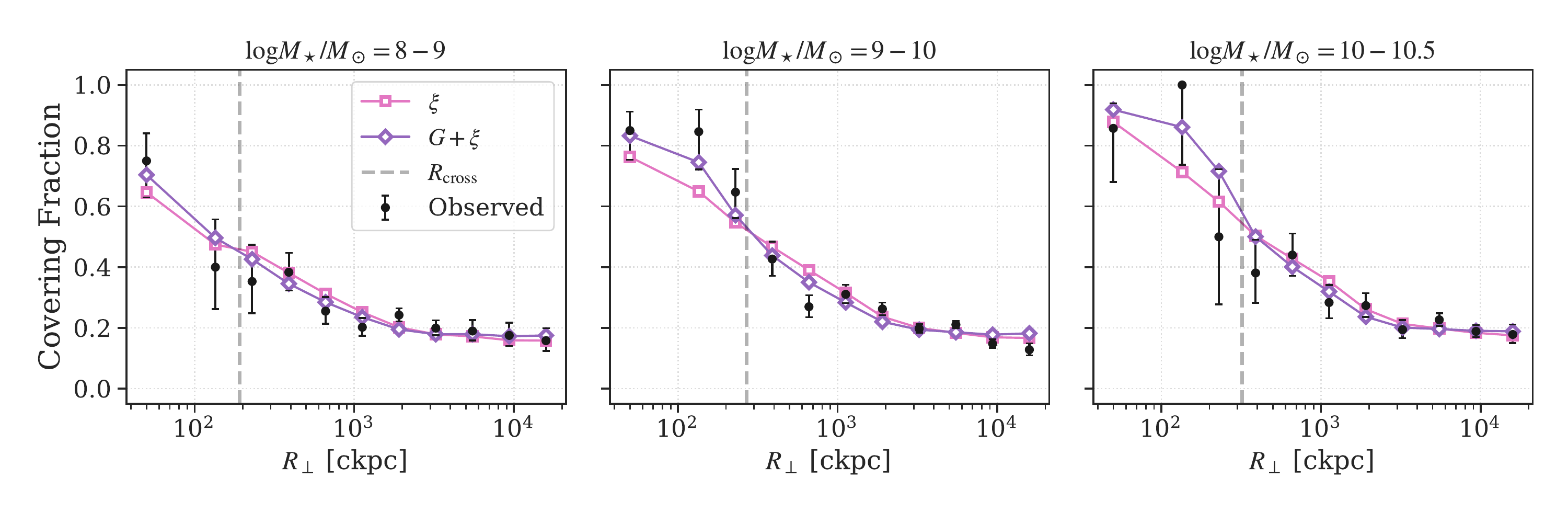}
\caption{Comparison of our two models to the empirical covering fraction as a function of impact parameter in comoving kpc in mass bins of $10^{8-9}M_{\odot}$, $10^{9-10}M_{\odot}$ and $10^{10-10.5} M_{\odot}$. The data are shown in black with 1$\sigma$ error bars. The single power-law model is shown in pink while the two-component model is shown in purple. The vertical dotted line denotes $R_{\rm cross}$ in each mass bin. Both models recreate the covering fraction of the data in all mass bins except for the lowest mass bin where the clustering signal disappears. The two-component model provides a better match to the data for galaxies of $M_{\star} > 10^{9}M_{\odot}$ at the lowest impact parameters where the single power law model underestimates the covering fraction. 
\label{fig:model_comparison}}
\end{figure*}

\subsubsection{Other Two-Component Models}
\label{section:othermodels}

While the single power-law clustering model does an adequate job reproducing the data on large spatial scales, its contribution is insufficient at $R_{\perp}$ $\lesssim$ 200 kpc as can be seen in Figure \ref{fig:model_comparison} (pink curve). Furthermore, the primary goal of our study is to find the boundary between the CGM and IGM, and thus including a CGM component is essential for this purpose. We explored several candidate functional forms for this CGM component.  

We first investigated a two-component model where each component is represented by a power law, inspired by the 1-halo and 2-halo terms that are used to model the clustering of galaxies. 
The 3D and projected forms of the two absorber-galaxy correlation functions are given by Equations~\ref{eq:xi} and \ref{eq:chiperp}, respectively, and the two-component correlation function is the sum of these parts.
We also considered a model where the two-component correlation function is, in 3D, the maximum of the two power laws. 
This is similar to our chosen model, but with an inner power law rather than an inner Gaussian profile.

To rise above the outer power law component at small radii, the inner power law has to be steeper. 
In practice, the two power law indices turned out to be similar, yielding essentially the same result as a single power law fit.
This outcome is not unexpected: the enhancement in the incidence rate or surface density of gas near galaxies often does not resemble a steepening power law at small radii \citep{zhu14,lan20}.

In those studies, the enhancement is better described by a function that declines gradually (compared to a power law) at small radii and quickly at large radii.
The top-hat function, which has amplitude $A$ inside a boundary and amplitude 0 outside the boundary, is an extreme example of this class.
Our adopted Gaussian profile allows a smoother transition between the CGM-like and outer components of the model.
However, we note that a fit to the data combining a inner 3D top-hat with an outer power law yields an $R_{\rm cross}$($M_{*})$ that is effectively indistinguishable from the one that emerges from the Gaussian component model.

\subsection{Model Comparison}
In addition to comparing the two models to each other, Figure \ref{fig:model_comparison} compares the models to the empirical covering fraction as a function of impact parameter and mass. The data are shown in black with 1$\sigma$ error bars. The single power-law model is shown in pink while the two component model is shown in purple. Both models recreate the covering fractions in all mass bins at all values of $R_{\perp}$ except for one data point in the $\log M_*/M_\odot = 9 - 10 $ bin at $R_{\perp}$ $\approx$ 200 kpc. Moreover, the two models make different predictions at low $R_{\perp}$ except for in the lowest mass bin ($\log M_*/M_\odot < 9 $) where there is no discernible excess above the clustering signal. This does not preclude the presence of a CGM  around these galaxies, but rather suggests that we require more data at lower $R_{\perp}$ for galaxies with  $\log M_*/M_\odot < 9 $ to be able to constrain $R_{\rm cross}$ at these masses. 

\begin{figure*}[ht!] 
\plotone{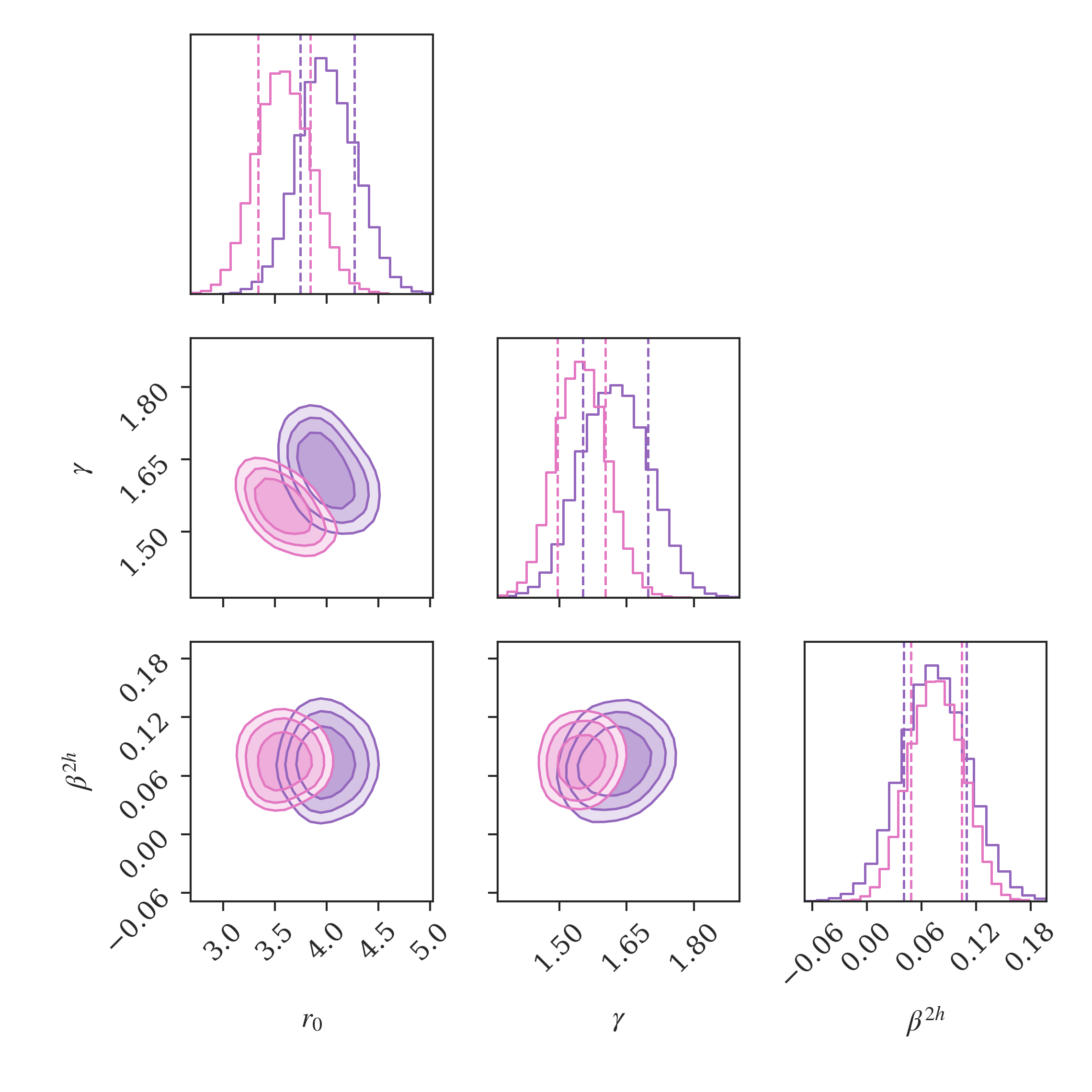}
\caption{Comparison of the two-halo 3D cross correlation posteriors between the two-component model ($r_0 = 3.99^{+0.28}_{-0.24}$ cMpc, $\gamma = 1.62 \pm 0.07$) and the single power-law model ($r_0 = 3.58^{+0.28}_{-0.24}$ cMpc, $\gamma = 1.55 \pm 0.05$). The two models are consistent with each other within the 1$\sigma$ limits and have a power-law slope consistent with the absorber-galaxy 3D cross correlation found in the literature \citep[e.g.][]{tejos14} of $\gamma = 1.7 \pm 0.1$.  
\label{fig:corner_comparison}}
\end{figure*}

The two-halo only model under-predicts the observed signal for galaxies at intermediate masses ($\log M_*/M_\odot = 9-10$).  The two component model does better for galaxies of $M_{\star} = 10^{9-10}M_{\odot}$ at the lowest impact parameters where the single power law model underestimates the covering fraction, although not significantly so. For  $R_{\rm cross} < 300$\,kpc, one detects 52~\ion{H}{1}\ systems where 46~systems are predicted. Assuming Poisson statistics, the two-halo only model is consistent with the data at 1$\sigma$ level. Analogous to the one-halo term of galaxy-galaxy clustering, the data themselves do not require an enhanced covering fraction of \ion{H}{1}\ absorption that we identify as the CGM.

We find the 1-halo component has a stronger clustering mass dependence, $\beta^{1h} \simeq 0.14 \pm 0.07$, than the two-halo term, $\beta^{2h} \simeq 0.08 \pm 0.03$. We also find the 2-halo clustering terms in each model to be internally consistent with each other as seen in Figure \ref{fig:corner_comparison}.

\section{Results}\label{section:results}
\subsection{Clustering Mass Dependence}

As seen in Figure \ref{fig:corner_2h}, we find the clustering parameters to be $r_0 = 3.6 \pm 0.3$ cMpc, $\gamma = 1.6 \pm 0.5$. $r_0$ and $\gamma$ are consistent with those found in \cite{tejos14} who find $r_0 = 3.7 \pm 0.1$ cMpc and $\gamma = 1.7 \pm 0.3$. We also find a mass dependence of the absorber-galaxy clustering of $\beta^{2h} = 0.07^{+0.3}_{-0.2}$. 

 We find the the two component model better fits the data as can be seen in Figure \ref{fig:model_comparison}. Specifically, the two component model better matches the covering fraction for galaxies of $M_{\star} > 10^{9-10}M_{\odot}$ at the lower impact parameters where the single power law model underestimates the covering fraction. In addition, we find the two-component model reproduces the mass dependence of the 2-halo clustering term, $\beta^{2h} \simeq 0.07$ while also producing a stronger mass dependence of the 1-halo clustering term, $\beta^{1h} \simeq 0.14$. 

\subsection{Physically-Motivated Extent of the CGM}

As mentioned above, using the two-component model produces an estimate of $R_{\rm cross}$, a natural metric for the extent of the CGM. This 3-D distance demarcates where the contribution to the clustering begins to be dominated by the CGM above the expected two-halo clustering due to isolated galaxy halos traced by \ion{H}{1}. $R_{\rm cross}$ can be viewed as the maximum radius to which an enhancement from the CGM could extend without over-predicting the data at large radii. 

\begin{figure}[t] 
\includegraphics[width= \linewidth]{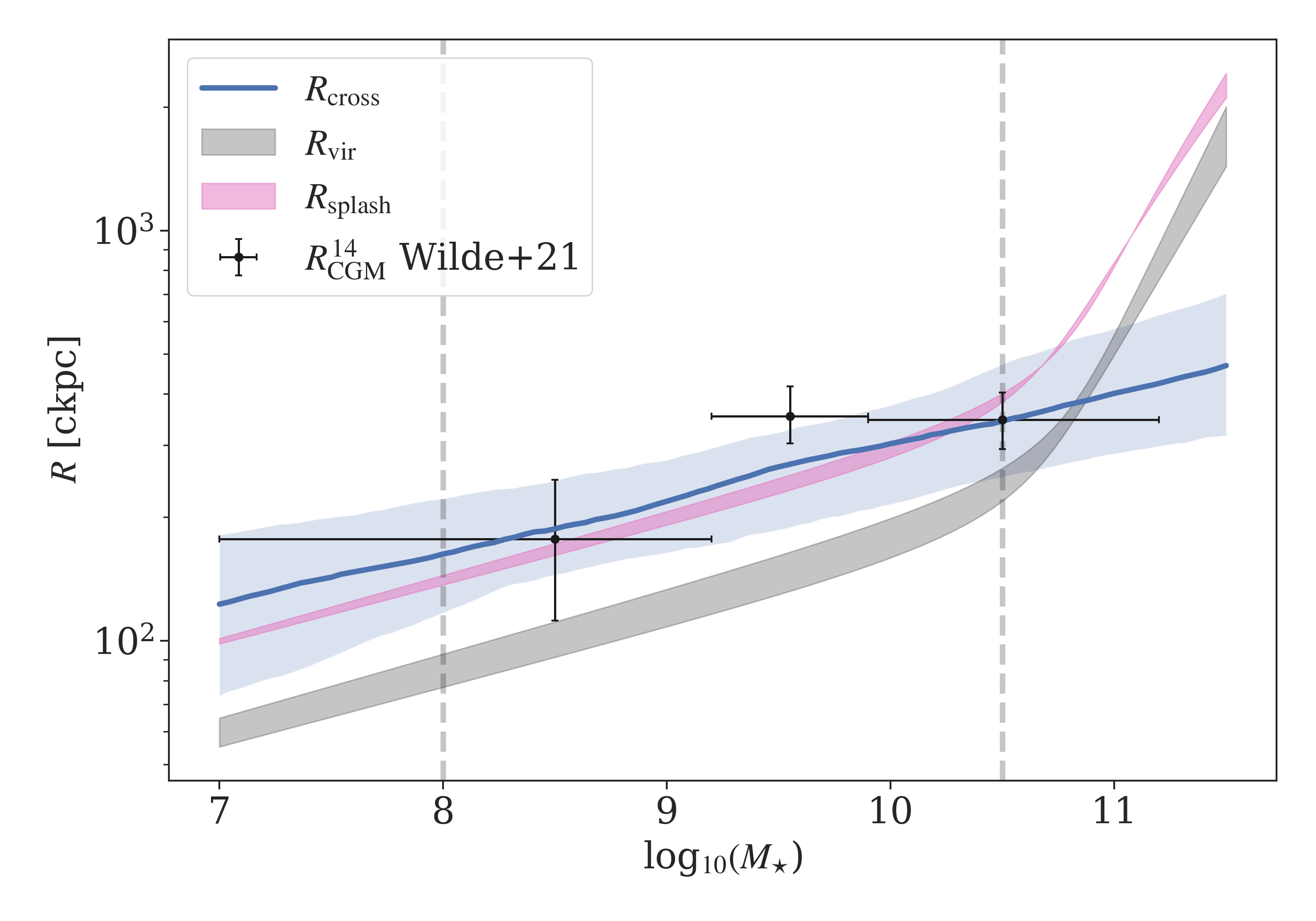}
\caption{A comparison of $R_{\rm cross}$ with the virial radius ($R_{\rm vir}$, grey filled region) as well as the splashback radius ($R_{\rm splash}$, pink shaded region) of the galaxy sample. The filled regions in $R_{\rm vir}$ and $R_{\rm splash}$ denote the redshift range for the galaxies in our sample ($0.1 \lesssim z \lesssim 0.48$).
The filled blue region represents the 1$\sigma$ limits of the distribution in $R_{\rm cross}$ while the blue line denotes the median of this distribution. The black crosses correspond to the values published in Paper I. The vertical dotted lines denote the mass range of $8 < \log(M_{\star}/M{\odot}) < 10.5$ to which we limited the fitting in our MCMC analysis in Figure \ref{fig:model_comparison}.   \label{fig:rcross}}
\end{figure}

In Figure \ref{fig:rcross}, we see $R_{\rm cross}$ (blue) compared with the spread in virial radii of the galaxy sample (grey filled region). The filled blue region represents the 1$\sigma$ limits of the distribution in $R_{\rm cross}$ while the blue line denotes the median of this distribution. We find $R_{\rm cross}$ is $\sim 2 \pm 0.6 R_{\rm vir}$ for galaxies in the range $8 < \log(M_{\star}/M{\odot}) < 10.5$. The black crosses correspond to the values published in Paper I defined as the extent where there is 50\%chance to see \ion{H}{1} absorption above $10^{14}$ cm$^{-2}$. The vertical dotted lines denote the mass range of $8 < \log(M_{\star}/M{\odot}) < 10.5$ that was used in our MCMC analysis. Above this range, we see a change in the relation of the virial radius with stellar mass, and below this mass range, we find little to no correlation between absorbers and galaxies. \ref{fig:model_comparison}). 

We also calculated the splashback radius, $R_{\rm sp}$, using the method from \cite{diemer18} and encoded in the COLOSSUS\footnote{https://bdiemer.bitbucket.io/colossus/} package. This radius denotes the location at which particles reach the apocenter of their first orbit. We find excellent agreement of $R_{\rm cross}$ with the results in Paper I and $R_{\rm cross}$ neatly matches the splashback radius for galaxies in this mass range. We discuss these results in more detail below.

\section{Discussion}\label{section:discussion}

Both of the models we investigate do an adequate job of recreating the cross correlation signal at all impact parameters and masses $10^{8} < M_{\star} < 10^{10.5} M_{\odot}$ as seen in Figure \ref{fig:model_comparison}.  It is not entirely clear that the single power law model has any physically-consistent meaning, however. Effectively, it would seem to signify that every time one measures \ion{H}{1} absorption at the same redshift  as a particular galaxy ($|\Delta v| <$ 500 km s$^{-1}$), the absorption is always due to {\emph{another galaxy's CGM}}. Note, we would conclude this for all galaxies, i.e. each has no CGM and only neighbors with a CGM. This is clearly impossible.
The two-halo-only model for the CGM effectively breaks down when
the galaxies lie within the halo under consideration, i.e. when they ``mix."  
We cannot and do not try to distinguish between the two. However, our formalism does allow one to identify the outer extent of this ``mixing."

The two-component model asserts that galaxies with $M_{\star}>10^{8} M_{\odot}$ have a CGM, an assumption that is motivated by previous survey results \citep[e.g.][]{werk13}. Additionally, this model is able to better recreate the data -- from the combined datasets of \cgmsq$+$ CASBaH, which together represent the largest sample of galaxies with confirmed spectroscopic redshifts in the foregrounds of UV-bright QSOs with high-resolution absorption spectroscopy  --  both at smaller impact parameters and at $M_{\star} > $ 10$^{9}$ $M_{\odot}$.  

The much larger number of galaxies at larger impact parameters drives the fit of the models to the data. There is, however, a $ > 1\sigma$ inconsistency between the two-halo only model and the data at $R_{\perp} \sim 200 $ and for both models at $R_{\perp} \sim 600$ in the log$M_{\star} =  9-10 M_{\odot}$ mass range. The latter inconsistency may be due to cosmic variance or the assumption that the absorber-galaxy measurements are independent and are not correlated, which would increase the scale of the error bars at $R_{\perp} \sim 600$. 

\subsection{Comparing the mass dependence of the single and two-component models}
Our galaxy sample includes a large number of galaxies at low ($<500$ kpc) impact parameters which allows us to better model the regime in which the two-halo galaxy clustering becomes dominated by the signal of galaxies that inhabit the same dark matter halo, the one-halo term. By separating these two terms in the manner presented here, we can disentangle the large scale clustering as well as the contribution of the CGM to the 3D correlation of absorbers and galaxies. 

Our analysis finds nearly identical terms for the mass dependence of the clustering at large scales, $\beta^{2h}$ as well as the contribution of absorbers at random, $C_0$ and $\alpha$. We do find a stronger mass dependence in the one-halo term, $\beta^{1h}$ than at larger scales. This can be seen in Figure \ref{fig:model_comparison} where the correlation steepens in higher mass bins.

\subsection{Absorber-Galaxy Bias}
Our covering fraction analyses provide an estimate of the galaxy-absorber correlation function, $\xi_{ag}$ (eq. \ref{eq:xi}). 
Here, we test if the mass dependence of $\xi_{ag}$ outside the CGM is consistent with absorption systems and galaxies simply being two independent tracers of the same underlying dark matter distribution. 
Assuming both tracers have linear bias, $\xi_{ag}$ should be equal to $b_a b_g \xi_{\rm DM}$, where $b_a$ and $b_g$ are the absorber and galaxy bias, respectively, and $\xi_{\rm DM}$ is the dark matter 3D correlation function.
Following \citet{tinker10} (hereafter, T10), we assume the dark matter correlation function can be described by a power-law function of radius with index $\gamma = 1.62$.
We fix the power-law index in the $\xi_{ag}$ determined by fitting a single power-law to the data to this same value, with which it is consistent.
With the above assumptions, $\xi_{ag} = (r/r_0(M))^{-\gamma} = b_a b_g \xi_{\rm DM}(r)$.
The radial dependence cancels, leaving the proportionality $r_0(M)^\gamma \propto b_a b_g$.

\begin{figure}[ht!] 
\includegraphics[width=\linewidth]{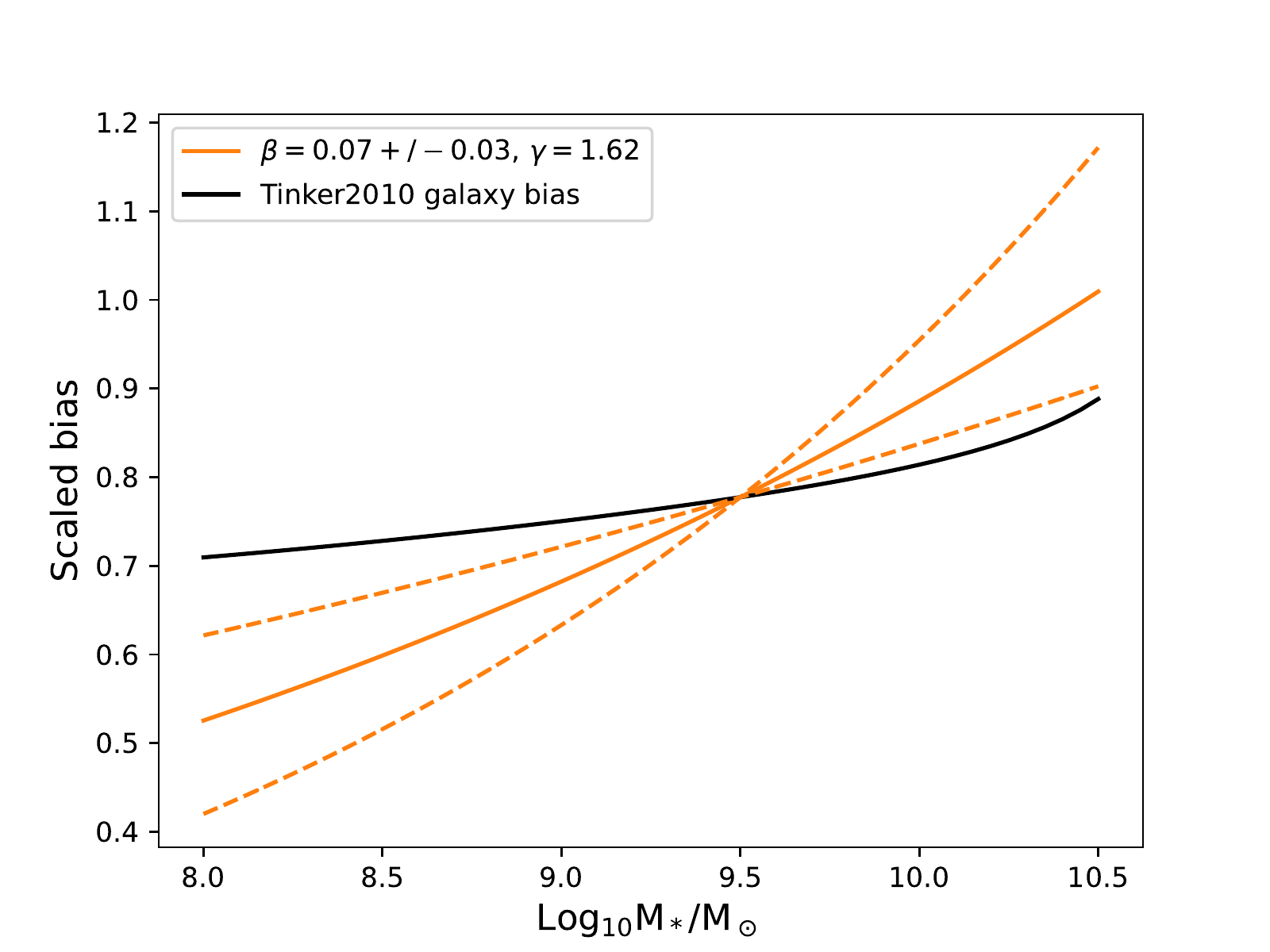}
\caption{A comparison of the slopes of the relative bias as a function of mass derived from our analysis (orange) compared to \cite{tinker10} (T10, black). The dashed lines correspond to the ranges spanned by the $1\sigma$ limits in in $\beta^{2h}$. The relative bias, $r_0(M) \propto (M_{\star}/ M_0)^{\gamma\beta}$, are normalized to the value of T10 at log $M_{\star}/M_\odot = 9.5$. We find a steeper mass dependence than T10, but the significance of the difference is less than $2\sigma$. \label{fig:bias_mass}}
\end{figure}

We show a scaled $r_0(M)^\gamma$ in Figure \ref{fig:bias_mass} along with the galaxy bias as a function of stellar mass from T10 and implemented in the \texttt{COLOSSUS} package \citep{diemer18}.
If $b_a$ is constant and the assumptions stated above hold, $r_0(M)^\gamma$ should have the same mass dependence as galaxy bias.
While there is a visually apparent difference between the galaxy bias and the best-fit $r_0(M)^\gamma$, this difference is not significant at a $2\sigma$ level and so is merely suggestive. 
If the difference is real, it could be a consequence of the \ion{H}{1}\ mass per dark matter mass being a function of overdensity.
Up to the overdensities at which $M_{star}=10^{10.5}$ $M_{\odot}$ galaxies tend to be found, this function would be increasing: \ion{H}{1}\ would be less common in low density regions than in higher density filaments.
This behavior would be consistent with theoretical expectations (e.g., \citealt{hui97, schaye01, dave10}) and observations (e.g., \citealt{rudie12b, burchett20}).

\subsection{Comparison to Previous Work}
One of the key aspects of this analysis is determining the mass dependence of the extent of the \NHI$>10^{14}$ cm$^{-2}$ for which our model provides a direct metric, $R_{\rm cross}(M_{\star})$. We compare our resulting $R_{\rm cross}(M_{\star})$ to the method and results from Paper I in Figure \ref{fig:rcross}. The result of Paper I, $R^{14}_{\rm CGM}$, which are based only on the \cgmsq survey are shown as black crosses in the mass bins they span in that paper. We also compare the method used in that paper to determine $R^{14}_{\rm CGM}$, the radius at which the probability of detecting \NHI $>10^{14}$ cm$^{-2}$ is $>50\%$, calculated with the two-component model using the combined \cgmsq$+$ CASBaH surveys and find it to be consistent within 1$\sigma$ with our newer model for $R_{\rm cross}(M_{\star})$. We find that our mass dependent estimate of the extent of the CGM, $R_{\rm cross}(M_{\star})$ corroborates the findings of Paper I that the \NHI$>10^{14}$ cm$^{-2}$ extends to approximately twice the virial radius ($\sim 2 \pm 0.6 R_{\rm vir}$).

One of the main strengths of the CGM$^2 + $ CASBaH sample is the large number of galaxies at small projected separations ($<$1 Mpc). This allows us to investigate the smaller scale regime in more detail within the context of similar studies such as \cite{tejos14} (hereafter, T14) who uses a single power law model to measure the two-point correlation between \ion{H}{1} and galaxies above \NHI $> 10^{14}$ cm$^{-2}$. In this work they break up their measurements into SF vs non-SF samples while we do not. Our sample however is dominated by the more common SF galaxies and we will compare our results to their SF sample. Comparing our cross-correlation results with T14, we find good agreement between the results in T14, $r_0^{\rm T14} = 3.8 \pm 0.2$ Mpc, $\gamma = 1.7 \pm 0.1$ and the results from both models presented here, $r_0 = 3.99^{+0.28}_{-0.24}$ Mpc, $\gamma = 1.62 \pm 0.07$) and the single power-law model ($r_0 = 3.58^{+0.28}_{-0.24}$ Mpc, $\gamma = 1.55 \pm 0.05$. We find a mass dependence of this cross-correlation, however as parameterized by $\beta^{2h}$.

Our results are slightly in tension with \cite{momose21} who find galaxies in the $10^{9-10} M_{\odot}$ range dominate their \ion{H}{1}-galaxy cross correlation signal. We find the largest mass bin sample to have the most elevated covering fractions at low impact parameter.

\subsection{Physical Extent of Galaxy Halos}
Astronomers often use the viral radius as a means to describe the characteristic size of galaxy halos and it is convenient to compare this to the extent of the gaseous galactic atmosphere as we have done here and in Paper I. The virial radius is typically defined in terms of the spherical overdensity mass definition which is based on the radius which encloses an overdensity of 200 times the critical or mean density, i.e., $R200_c$ and $R200_m$. Because the mean and critical densities are decreasing over cosmic time, this can lead to a pseudo-evolution as pointed out in \cite{diemer13}. In addition, subhalos show evidence of being stripped outside the virial radius of clusters \citep{behroozi14}. 

An alternative physically motivated halo scale is the splashback radius, $R_{\rm sp}$ \citep[][]{diemer14, adhikari14, more15}.  This radius effectively distinguishes infalling material from matter orbiting in the halo. We compare our results to the splashback radius in Figure \ref{fig:rcross} and find that our estimate of the extent of the \ion{H}{1} CGM, $R_{\rm cross}$, neatly aligns with $R_{\rm sp}$ over the mass range $10^8 < M_{\star}/M_{\odot} < 10^{10.5}$. This result implies that \Rsp\ is a better approximation of the CGM extent than the more commonly used viral radius. 

\cite{oniel21} compared \Rsp\ as estimated from dark matter and gas profiles in the IllustrisTNG simulations and found that the gas \Rsp\ is consistently smaller than the dark matter \Rsp. However, they were looking at much more massive halos $M_{\rm halo} > 10^{13}$ in which shocks dominate the gas distribution. Nonetheless, the fact that  $R_{\rm cross} \approx R_{\rm sp}$ at the mass ranges considered here ($M_{\rm halo} ~ 10^{10-12} M_{\odot}$) is intriguing. The halo mass accretion rate generally sets whether $R_{\rm sp}$ exceeds $R_{\rm vir}$; a rapid accretion rate will impact the growth of the gravitational potential well, leading to $R_{\rm sp} < R_{\rm vir}$. If the location of $R_{\rm cross}$ reflects the extent of orbiting gas in a halo, then our observational results imply a halo mass accretion rate that is slow enough to keep the apocenters of orbiting structures at large radii.

Another way of defining the extent of the CGM is to use the boundary of the pressure-supported CGM. For galaxies with halo masses $ \gtrsim 10^{11.5} M_{\odot}$ ($M_{\star} \approx 10^{9.8} M_{\odot}$), this pressure support comes from fact that the gas that has fallen into the gravitational potential well is virially shocked and cannot cool within a Hubble time \citep[]{binney77, rees77, silk77}. For the galaxies in our survey, which are predominately below this halo mass, however, the  gas would rapidly cool and thus this pressure support might come from galactic winds. \citet{fielding17} and \citet{lochhaas18} show that supernovae winds with reasonable mass loading efficiencies could shock the gas to distances past the virial radius and account for the survival of cool gas at these large radii. Using a more comprehensive model of the multiphase CGM, \cite{fielding22} show that SF in the galactic disk can slow cooling and accretion as part of a global preventive self-regulation mechanism. In addition, the winds can transport cold clouds to large radii, consistent with these constraints from our combined survey data. 

\section{Summary}\label{section:summary}

Herein, we have examined the associations of galaxies with \lya absorption $z < 0.48$ to explore the spatial profile of this gas and the mass dependence of the profile. Specifically, we have combined the \cgmsq and CASBaH \ion{H}{1} measurement and constructed a catalog of 7244 absorber-galaxy pairs around 28 QSO sightlines (6589 absorber-galaxy pairs when we restrict our galaxy sample to galaxies with $8 < \log M_{\star}/M_{\odot} < 10.5$).  The \cgmsq survey has better sampling of galaxies at low impact parameter while CASBaH samples galaxies out to 20 cMpc. This allows us to characterize the \ion{H}{1}\ profile via the covering fraction as a tracer of the gas.

\begin{enumerate}
    \item By modeling the covering fraction as a power-law with a mass dependent length scale, we find good agreement with previous studies, such as T14, of our clustering amplitude and power law slope parameters.
    
    \item In Section \ref{sec:single-pl}, we find the clustering scale has a mass dependence with a power-law slope of $\beta^{2h} = 0.08 \pm 0.03$.
    
    \item We compare the slope of our absorber-galaxy bias to the galaxy-dark matter bias of \citet{tinker10}. The absorber-galaxy bias is a steeper function of galaxy mass than the galaxy-dark matter bias. However, this difference is only significant at a sub-$2\sigma$ level.

    \item We model the data with an exclusionary two-component model where we adopt an inner-CGM Gaussian profile to describe the data at smaller impact parameters and the customary two-halo single power-law model at larger impact parameters. This model faithfully reproduces the data for galaxies $M_{\star} > 10^8 M_{\odot}$. 
    
    \item The two component model allows us to calculate the crossover radius, $R_{\rm cross}(M_{\star})$, where the models are equal. $R_{\rm cross}(M_{\star})$ represents a soft upper estimate of the furthest impact parameter needed to optimally fit the inner CGM component.  We then use $R_{\rm cross}$ as an estimate of the extent of the CGM and find $R_{\rm cross}(M_{\star}) \approx 2 \pm 0.6 R_{\rm vir} $ for galaxies $10^8 \leq M_{\star}/M_{\odot} \leq 10^{10.5}$. Additionally, we find excellent agreement between $R_{\rm cross}(M_{\star})$ and the splashback radius, $R_{\rm sp}$ for galaxies in this mass range.

\end{enumerate}
 
\section{Acknowledgments}
MCW, KT, and JKW acknowledge support for this work from NSF-AST 1812521, NSF-CAREER 2044303, the Research Corporation for Science Advancement, grant ID number 26842. Support for the CASBaH HST programs HST-GO-11741 and HST-GO-13846 was provided
through grants from the Space Telescope Science Institute under NASA contract NAS5-26555.  

Support for the CASBaH HST programs HST-GO-11741 and HST-GO-13846 was provided through grants from the Space Telescope Science Institute under
NASA contract NAS5-26555.

The CGM$^{2}$ Survey would not have been possible without the substantial contributions from a dedicated group of UW undergraduate Student Quasar Absorption Diagnosticians, the Werk SQuAD, with over 50 individual undergraduate research assistants since 2016. The SQuAD confirmed all auto-fitted galaxy spectroscopic redshifts by eye, identified absorption systems along every quasar line of sight, and measured various spectroscopic properties (e.g. SFRs) of the nearly 1000 galaxies included in the survey. We are deeply grateful to work with such motivated and enthusiastic students. 

\bibliography{main.bbl}
\end{document}